\documentclass[runningheads]{llncs}

\usepackage{microtype}

\usepackage{booktabs}
\usepackage{cite} %
\usepackage{enumitem}
\usepackage{mathtools}
\usepackage{tabularx}
\usepackage{xcolor}

\usepackage[T1]{fontenc}

\usepackage{tikz}
\usetikzlibrary{fit, patterns, shadows}

\usepackage{pgfplots}
\pgfplotsset{compat=1.17}
\usepgfplotslibrary{colorbrewer}
\usepgfplotslibrary{groupplots}

\usepackage{hyphenat}
\usepackage{tcolorbox}
\tcbuselibrary{breakable}

\usepackage{environ}

\let\paragraph\subsubsection

\usepackage{float}
\floatstyle{ruled}
\newfloat{listing}{tbp}{lop} %
\floatname{listing}{Listing}

\usepackage{listings}

\def\codefontsize{\small} %

\newlength{\PyXLeftMargin}
\newcommand{\lstinput}[1]{%
  \lstinputlisting{listings/#1.py}%
  \raggedleft\footnotesize\texttt{listings/style.tex} missing; use \texttt{lstlistings} for highlighting
}

\newcommand{\pyOp}[1]{\texttt{#1}}
\newcommand{\pyStr}[1]{\texttt{"#1"}}%

\newlength{\FvXLeftMargin}
\usepackage{fancyvrb,xcolor}
\newcommand{\useFancyVrb}{
  \input{listings/style.tex}%
  \renewcommand{\lstinput}[1]{%
    \bgroup{}%
    \fvset{
      numbers=left,
      numbersep=6pt,
      xleftmargin=\FvXLeftMargin,
      commandchars=€,
      vspace=0pt,
    }%
    \renewcommand{\theFancyVerbLine}{%
      \color{gray}\codefontsize\arabic{FancyVerbLine}%
    }%
    \codefontsize\ttfamily%
    \input{listings/##1.tex}%
    \egroup{}%
  }%
  \renewcommand{\pyOp}[1]{{\ttfamily\PY{o}{##1}}}%
  \renewcommand{\pyStr}[1]{{\ttfamily\PY{l+s+s2}{\PYZdq{}}\PY{l+s+s2}{##1}\PY{l+s+s2}{\PYZdq{}}}}%
}

\IfFileExists{listings/style.tex}{\useFancyVrb{}}{}

\usepackage[hidelinks, pdftex]{hyperref}

\hypersetup{bookmarksdepth=subsection}

\begin{document}

\title{Automated Selfish Mining Analysis for DAG-Based PoW Consensus Protocols}
\titlerunning{Automated Selfish Mining Analysis}
\author{Patrik Keller}
\authorrunning{P. Keller}
\institute{University of Innsbruck, Austria\\\email{patrik.keller@student.uibk.ac.at}}
\maketitle

\begin{abstract}
  Selfish mining is strategic rule-breaking to maximize rewards in proof-of-work protocols~\cite{eyal2014MajorityNot}.
  Markov Decision Processes (MDPs) are the preferred tool for finding optimal strategies in Bitcoin~\cite{sapirshtein2016OptimalSelfish, gervais2016SecurityPerformance} and similar linear chain protocols~\cite{zhang2019LayCommon}.
  Protocols increasingly adopt DAG-based chain structures~\cite{wang2023SoKDAGbased}, for which MDP analysis is more involved~\cite{barzur2020EfficientMDP}.
  To date, researchers have tailored specific MDPs for each protocol~\cite{sapirshtein2016OptimalSelfish, gervais2016SecurityPerformance, zhang2019LayCommon, barzur2020EfficientMDP, hou2021SquirRLAutomating, keller2023TailstormSecure}.
  Protocol design suffers long feedback loops, as each protocol change implies manual work on the MDP.
  To overcome this, we propose a generic attack model that covers a wide range of protocols, including Ethereum Proof-of-Work, GhostDAG~\cite{sompolinsky2021PhantomGhostdag}, and Parallel Proof-of-Work~\cite{keller2022ParallelProofofwork}. %
  Our approach is modular: we specify each protocol as a concise program, and our tooling then derives and solves the selfish mining MDP automatically.

\end{abstract}

\section{Introduction}

In the context of proof-of-work cryptocurrencies, ``selfish mining'' entails the deliberate deviation from the protocol to maximize personal rewards, e.g., by temporarily withholding blocks.
The first selfish mining strategy was proposed by Eyal and Sirer in 2014 against Bitcoin~\cite{eyal2014MajorityNot}.
Soon after, other researchers started to employ Markov Decision Processes (MDPs) to identify and evaluate optimal behavior~\cite{sapirshtein2016OptimalSelfish, gervais2016SecurityPerformance}.
Using these strategies, attackers controlling more than a third of the system's hash rate can reap excessive rewards.
Depending on the communication assumptions, the threshold is even lower.

In 2019, Zhang and Preneel~\cite{zhang2019LayCommon} comprehensively mapped and analyzed various defense mechanisms. %
Their findings indicate that all designs for improving incentive compatibility compromise at least one of the other security properties.
Many recent protocol attempt to close this gap by adopting a directed acyclic graph (DAG) structure~\cite{sompolinsky2021PhantomGhostdag, keller2022ParallelProofofwork, keller2023TailstormSecure, pass2017FruitChainsFair, abraham2023ColordagIncentivecompatible}. %
But MDP-modeling requires enumerating all feasible states, and doing this for arbitrary DAGs is not tractable.
Hence, the DAG-based protocols have, so far, escaped MDP-based selfish mining analysis.

Fortunately, recent research has provided new options for analysing complex protocols.
Probabilistic Termination%
~\cite{barzur2020EfficientMDP} enables the \emph{exact} solution of selfish mining MDPs with a tenfold efficiency improvement.
When this approach meets its limits, one can turn to \emph{approximating} reinforcement learning techniques%
\cite{hou2021SquirRLAutomating, keller2023TailstormSecure, barzur2022WeRLmanTackle}, which can handle arbitrarily large state spaces, including infinite ones~\cite{sutton2018ReinforcementLearning}.

The remaining bottleneck arises from tailoring the MDPs.
So far, this involved a lot of manual work, basically starting from scratch for each protocol~\cite{sapirshtein2016OptimalSelfish, gervais2016SecurityPerformance, zhang2019LayCommon, barzur2020EfficientMDP, hou2021SquirRLAutomating, keller2023TailstormSecure}.
To overcome this, we propose a generic attack model that covers a wide range of DAG protocols.
We specify each protocol as a concise program.
Our tooling then generates the selfish mining MDP automatically.

We demonstrate our tooling by specifying and analyzing multiple protocols.
Bitcoin allows to validate our approach against existing results.
Ethereum Proof-of-Work and its Byzantium update serve as examples for (mostly) linear-chain protocols that carefully introduce DAG structure to adjust the reward scheme and chain preference rule.
GhostDAG~\cite{sompolinsky2021PhantomGhostdag} and Parallel Proof-of-Work~\cite{keller2022ParallelProofofwork} represent the newer class of DAG-protocols that, so far, has been considered out of reach for MDP-based analysis.
Notable results are that the optimal selfish mining policies against Ethereum, Byzantium, and GhostDAG involve creating more than one fork, and that relatively weak miners can benefit from selfish mining in GhostDAG.
The most resilient protocol is Parallel Proof-of-Work.

We proceed as follows:
Section~\ref{sec:background:mdp} provides relevant background information on MDPs;
Section~\ref{sec:attack_space} enumerates our assumptions and sketches the approach;
Section~\ref{sec:spec} explains how we specify protocols;
Section~\ref{sec:mdp} documents how we search optimal selfish mining policies in automatically derived MDPs;
Section~\ref{sec:optimizations} introduces two optimizations reducing the size of the state space;
Section~\ref{sec:results} documents all experiments and presents the results;
Section~\ref{sec:discussion} discusses our work and concludes.

\section{Markov Decision Processes (MDP)}\label{sec:background:mdp}

Finding incentive attacks against proof-of-work consensus protocols involves sequential decision-making, where an attacker must make a series of choices over time.
The outcomes of these decisions are non-deterministic due to the inherently stochastic nature of mining.
We use Markov Decision Processes (MDP) to model the problem and find the most profitable behavior.

An MDP is a probabilistic state machine. The \emph{agent}, observes the current state of the system and chooses an \emph{action}.
The MDP defines probabilistic state transitions, specifying the probability distribution for the next state based on the current state and the chosen action.
In our case, the agent models the attackers.
The defenders are not agents, they react passively to the actions of the attacker.

A sequence of state transitions, starting from the (probabilistic) initial state until the process ends, is called an \emph{episode}.
Each state transition may have an effect, such as assigning a reward. The agent's goal is to maximize a function of the effects (objective function), in our case, expected mining rewards per time.

The strategy an agent uses to decide which action to take in each state is called a \emph{policy}.
Solving an MDP is about finding a policy that maximizes the objective function.
For us, the optimal policy is the worst-case attack.

A good, textbook-sized introduction to the topic is provided by Sutton and Barto~\cite{sutton2018ReinforcementLearning}. The book offers a formal mathematical description of MDPs and covers solving algorithms from first principles. These methods can be broadly categorized as follows:

\begin{description}
    \item[Linear programming:] The Bellman equations provide an optimality condition for the policy.
      This induces a linear programming problem which can be solved exactly.
    \item[Dynamic programming:] Algorithms such as value iteration and policy iteration use the Bellman equations to define an iterative optimization step. Repeatedly applying this step converges to the optimal policy. When problem sizes become too large for linear programming, these methods offer practical solutions without losing much accuracy.
    \item[Approximate methods:] Modern reinforcement learning (RL) techniques, such as Q-learning or policy gradient methods, train a simplified surrogate model by observing the probabilistic state machine. The resulting policies may not be optimal in the original MDP, but these methods are highly versatile: they are applicable when exact solutions are infeasible and effective even when the underlying MDP is unknown or changes over time.
\end{description}

Dynamic programming has been successfully applied to Bitcoin \cite{sapirshtein2016OptimalSelfish, gervais2016SecurityPerformance,barzur2020EfficientMDP} and Ethereum Proof-of-Work~\cite{barzur2020EfficientMDP}.
Approximating methods have been applied to more complex protocols~\cite{hou2021SquirRLAutomating} or objective functions~\cite{barzur2022WeRLmanTackle}.

All these have in common that the researchers manually craft one MDP per problem and protocol.
Our goal is to derive MDPs automatically from the protocol specification.
We will use dynamic programming to provide optimal baseline results for small problem instances.
Applying RL to solve bigger problem instances approximately is an option for future work.

\section{Protocol-Generic Attack Space}\label{sec:attack_space}

We propose a protocol-generic state and action space (attack space) that is tractable for MDP-based analysis while still covering a useful range of attacks.

\paragraph*{Assumptions}
We achieve tractability by exploiting the following assumptions.

\begin{enumerate}[label=\textbf{A\arabic*.}, ref=\textbf{A\arabic*}, leftmargin=2.2em,]
  \item \label{ass:comm} \label{ass:first}
    Participants (miners) exchange blocks via reliable broadcast, and this is the only form of communication.
  \item \label{ass:dag}
    Blocks reference other blocks by their content, inducing a directed acyclic graph (BlockDAG) with a single root block (genesis).\label{ass:genesis}
  \item \label{ass:pow}
    Appending a new block to the DAG requires a proof-of-work.
    There is only one type of proof-of-work and the solving time is exponentially distributed.
  \item \label{ass:topo}
    The broadcast primitive delivers the blocks in topological order, i.e., a block is delivered only after all its parents have been delivered.
  \item \label{ass:det}
    Honest behavior follows deterministically from the structure of the BlockDAG and the order in which the individual blocks were delivered.
  \item \label{ass:consensus}
    We focus on consensus protocols:
    in ideal conditions, excluding attacks and communication delays, all miners agree on an ordering of (a subset of) the blocks which we call \emph{linear history}.
  \item \label{ass:attacker}
    We model all dishonest participants as a single agent, the \emph{attacker}.
  \item \label{ass:actions}
    The attacker may withhold and ignore blocks but otherwise behaves honestly.
  \item \label{ass:daa} \label{ass:last}
    We analyse selfish mining in the equilibrium state where the difficulty adjustment algorithm (DAA) has fully adjusted to the available hash rate and the attacker's policy.
\end{enumerate}

\begin{figure}
  \centering
  \pgfdeclarelayer{pattern}
  \pgfdeclarelayer{arrowbg}
  \pgfdeclarelayer{blockbg}
  \pgfdeclarelayer{arrow}
  \pgfsetlayers{pattern, arrowbg, arrow, blockbg, main}
  \tikzstyle{withheld}=[pattern=grid, pattern color=Set1-A!50]
  \tikzstyle{ignored}=[pattern=north west lines, pattern color=Set1-B!60]
  \tikzstyle{visible}=[pattern=north east lines, pattern color=Set1-C!40]
  \begin{tikzpicture}[x=2cm]
    \ifx\theblockid\undefined
      \newcounter{blockid}
    \else
      \setcounter{blockid}{0}
    \fi
    \newcommand{\block}[3]{
      \def\where{#1}
      \def\miner{#2}
      \def\edges{#3}
      \def\label{$\theblockid \mid \miner$}

      \node[draw, fill=white, rounded corners=2pt] at (\where) {\label};
      \begin{pgfonlayer}{blockbg}
        \node[draw, white, ultra thick, rounded corners=2pt] (b\theblockid) at (\where) {\label};
      \end{pgfonlayer}

      \begin{pgfonlayer}{arrow}
        \draw[->, >=stealth] (b\theblockid) \edges;
      \end{pgfonlayer}

      \begin{pgfonlayer}{arrowbg}
        \draw[->, ultra thick, >=stealth, shorten >=-1.5pt, white] (b\theblockid) \edges;
      \end{pgfonlayer}

      \stepcounter{blockid}
    }

    \begin{scope}[drop shadow={fill=white, shadow scale=1.1}]
      \block{0, 1}{g}{} %
      \block{1, 1}{d}{edge (b0)} %
      \block{1, 2}{a}{edge (b0)} %
      \block{2, 1}{d}{edge (b1)} %
      \block{2, 0}{a}{edge (b1)} %
      \block{3, 1}{d}{edge (b3) edge (b2)} %
      \block{3, 2}{a}{edge (b4) edge (b2)} %
      \block{4, 2}{a}{edge (b6)} %
      \block{4, 1}{d}{edge (b5) edge (b4)} %
    \end{scope}

    \begin{pgfonlayer}{pattern}
      \node[rounded corners, withheld, fit=(b6) (b7)] {};

      \node[rounded corners, ignored, fit=(b3) (b5) (b8)] {};

      \node[name=visible0, fit=(b0) (b1) (b2) (b3)] {};
      \node[name=visible1, fit=(b0) (b5) (b4)] {};
      \path[rounded corners, visible]
        (visible0.south east) rectangle (visible0.north west)
        (visible1.south east) rectangle (visible1.north west);
    \end{pgfonlayer}
  \end{tikzpicture}

  \tikzstyle{area}=[minimum width=4.3ex, minimum height=2.5ex, rounded corners=3pt]
  \begin{tikzpicture}
    \def\colsep{1em}
    \matrix[row sep=-4pt] {
      \node[right, font=\bfseries] {Blocks:}; &[\colsep]
        \node[]{$g$\strut}; & \node[right] {genesis\strut}; &[\colsep]
          \node[]{$d$\strut}; & \node[right] {defender\strut}; &[\colsep]
            \node[]{$a$\strut}; & \node[right] {attacker\strut}; \\
      \node[right, font=\bfseries] {Masks:}; &[\colsep]
        \node[area, withheld]{}; & \node[right] {withheld\strut}; &[\colsep]
          \node[area, ignored]{}; & \node[right] {ignored\strut}; &[\colsep]
            \node[area, visible]{}; & \node[right] {visible\strut}; \\
      };
  \end{tikzpicture}
  \caption{
    Example state.
    Vertices are blocks and arrows indicate parents.
    The available actions are \texttt{Release(6)}, \texttt{Consider(3)}, and \texttt{Continue}.
    Block 8 was just mined; it will become visible during the next \texttt{Continue} action.
  }
  \label{fig:state}
\end{figure}
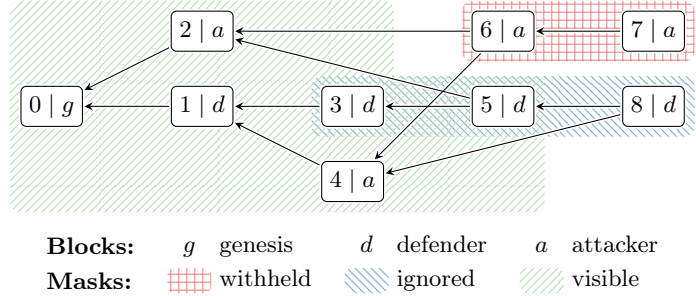

\paragraph*{State Space}
The assumptions~\ref{ass:comm}--\ref{ass:pow} and \ref{ass:attacker}--\ref{ass:daa} allow us to track the relevant state in a few variables:
\begin{enumerate}
  \item An adjacency list, describing the structure of the BlockDAG.
  \item A vector describing which blocks where mined by whom.
  \item An ignorance mask, describing which blocks are ignored by the attacker.
  \item A withholding mask, describing which blocks are withheld by the attacker.
  \item A visibility mask, describing which blocks have been delivered to the defender.
  \item Protocol-specific local states of the defender and the attacker.
\end{enumerate}

We do not consider the full cross product of these variables, as most combinations are invalid.
Due to topological delivery (\ref{ass:topo}), a block can become visible only after all parents are visible. Similarly, a block must remain ignored while any parent is ignored.
Naturally, the attacker can only withhold their own blocks.
As the blocks are delivered to the defender in topological order anyways, we apply the topological constraints to the withholding mask as well.
Additionally, the protocol-specific local states follow deterministically from the stepwise modification of the other variables (\ref{ass:det}).

Figure~\ref{fig:state} shows an example state.
We visualize the BlockDAG and the three masks for ignoring, withholding, and visibility.
We omit the miners' local states.
Figure~\ref{fig:state_btc} in Appendix~\ref{apx:traditional} shows another example with our attack space instantiated for the Bitcoin protocol.

\paragraph*{Action Space}
The original selfish mining MDP for Bitcoin uses only four actions (see Appendix~\ref{apx:traditional}).
We here use an (a-priori) unbounded number of actions and group them as follows.
\begin{description}
  \item[Consider(b)] to stop ignoring block b.
  \item[Release(b)] to stop withholding block b.
  \item[Continue] to wait until the next block is mined.
\end{description}
Induced by the topological restrictions of the state space, only of subset of actions is available at a time.

\paragraph*{Bounding the State and Action Spaces}
Exact MDP solving techniques work only if the number of states is finite.
This is clearly not the case, as the BlockDAG grows continuously.
We address this by truncating parts of the BlockDAG that are irrelevant for future protocol execution:
stale blocks and the prefix of the linear history that all miners already agree upon (common history).
Moreover, we define a limit on the number of blocks in the BlockDAG.
In states that meet this limit, we force the attacker to behave honestly taking away the options to withhold and ignore blocks.
By the consensus assumption \ref{ass:consensus}, the miners will agree on a linear history.
The common history will be truncated and the policy loops back to a smaller state.
This bounds the number of actions as well.

\paragraph*{Discussion of the Assumptions}
Most of the assumptions~\ref{ass:first}--\ref{ass:last} are uncontroversial and backed by related work.
We here give special attention to the last three and defer the rest of our arguments to Appendix \ref{apx:assumptions}.

Assumption~\ref{ass:attacker} restricts our model to a single attacking agent.
This is conservative, as it allows dishonest participants to collude and coordinate their attack.
This assumption is common practice since defenders typically cannot rule out collusion among the attackers~\cite{garay2020SoKConsensus}.
To address scenarios where attackers do not collude, our model can be expanded for multiple agents.
We think the main challenge here lies in defining a suitable scenario.
In the most obvious case where all participants act rationally, security collapses entirely~\cite{ford2019RationalitySelfdefeating}.

In Assumption~\ref{ass:actions}, we force the attacker to base their attack on ignoring and withholding blocks alone.
We not think this is overly restrictive:

First, as we lay out in Appendix~\ref{apx:traditional}, our attack space fully covers selfish mining against Bitcoin. We confirm this numerically in Section~\ref{sec:validation}.

Second, we deem it necessary from a practical point of view.
It allows us to derive the structure of the BlockDAG from honest behavior.
The alternative approach, starting from arbitrary blocks, checking them against the protocol's block validity rule, and then enumerating all valid choices, is not tractable.

Third, using local information alone, withholding is indistinguishable from benign network delays.
In the absence of network delays and withholding, local knowledge is global knowledge and consensus is easily solved.
We think, dealing with withholding is the key aspect of proof-of-work consensus.

Assumption~\ref{ass:daa} encapsulates the details of difficulty adjustment.
The DAA observes timestamps in the linear history and, in response, adjusts the difficulty of the proof-of-work puzzle.
The goal is to maintain a (roughly) constant growth rate for the linear history, even though the available hash rate changes over time.
This creates a feedback loop from the result of consensus (linear history, time) to the communication primitive (rate-limited by proof-of-work).

Expectedly, the attacker's actions affect the difficulty in the long term.
In fact, selfish mining in Bitcoin becomes profitable (expected reward per time) only after the DAA has adjusted to the policy.

We are aware that the DAA can be gamed, e.g., by faking time stamps~\cite{yaish2023UncleMaker} or tactically leaving and re-joining the system~\cite{negy2020SelfishMining, yaish2022BlockchainStretching}.
Following related work~\cite{eyal2014MajorityNot, sapirshtein2016OptimalSelfish, barzur2020EfficientMDP}, we keep our model tractable by assuming the DAA is in a stable state.

\section{Specification of Protocols}\label{sec:spec}

\begin{figure}[tbp]
  \centering
  \begin{tikzpicture}[>=stealth]

    \newcommand\Section[3]{
      \node[align=center, font=\bfseries] (a) {\strut#2\strut};
      \node[align=center, anchor=north, font=\ttfamily] at (a.south) (b) {\strut#3};
      \node[fit=(a) (b), draw, rounded corners, dashed, outer sep=1pt] (#1) {};
    }

    \begin{scope}

      \begin{scope}[xshift=-.106\linewidth]
        \Section{wa}{write\\access}{append(*)}
      \end{scope}

      \begin{scope}[xshift=.106\linewidth]
        \Section{ra}{read\\access}{parents(b)\\children(b)\\miner\_of(b)\\genesis\\G}
      \end{scope}

      \node[fit=(wa) (ra)] (bdi) {};

      \node[font=\large\bfseries] (bdl) at ([yshift = 5ex] bdi.north) {BlockDAG\strut};

      \node[fit=(wa) (ra) (bdl), draw, rounded corners, outer sep=1pt, inner sep=4pt] (bd) {};

    \end{scope}

    \begin{scope}[xshift=.55\linewidth]

      \begin{scope}[xshift=-.11\linewidth]
        \Section{si}{simulation\\interface}{init()\\update(b)\\mining()}
      \end{scope}

      \begin{scope}[xshift=.11\linewidth]
        \Section{oi}{observation\\interface}{history()\\coinbase(b)\\progress(b)}
      \end{scope}

      \node[fit=(si) (oi)] (psi) {};

      \node[font=\large\bfseries] (psl) at ([yshift = 5ex] psi.north) {Protocol Specification\strut};
      \node[font=\bfseries, below=-4pt] at (psl.south) {``miner''};

      \node[font=\bfseries] (ls) at ([yshift=-7ex] psi.south) {local state};

      \draw[->] (si) -- node[anchor=east] {modifies} (ls);
      \draw[->] (oi) -- node[anchor=west] {reads} (ls);

      \node[fit=(si) (oi) (psl) (ls), draw, rounded corners, outer sep=1pt, inner sep=4pt] (ps) {};

    \end{scope}

    \node[font=\bfseries] (pd) at ([yshift=-8ex] ps.south) {Protocol Designer\strut};
    \node[font=\bfseries] (e) at (bd.south |- pd) {Environment\strut};

    \node[fit=(bd.north east) (ps.south west)] (sep) {};
    \def\psicolor{Set1-A} %
    \draw[\psicolor, rounded corners=2pt, pattern=north east lines, pattern color=\psicolor]
      ([xshift=-10pt] sep.north) rectangle ([xshift=-2pt, yshift=4pt] sep.south |- e.south);
    \node[font=\bfseries, above, \psicolor] at ([xshift=-7pt] sep.north) {Protocol Specification Interface};

    \draw[->] (pd) -- node[anchor=west] {implements} (ps);

    \draw[ultra thick, white] (e) to[bend left=10] (wa); %
    \draw[->] (e) to[bend left=10] node[pos=0.25, anchor=east] {has} (wa);

    \draw[ultra thick, white] (e) to[bend right=10] (ra); %
    \draw[->] (e) to[bend right=10] node[pos=0.5, anchor=west] {has} (ra);

    \draw[ultra thick, white] (e) to[bend right=10] (ps); %
    \draw[->] (e) to[bend right=10] node[pos=0.35, anchor=north] {calls} (ps);

    \draw[ultra thick, white] (si) -- (ra.east |- si); %
    \draw[->] (si) -- node[pos=0.33, anchor=south] {has} (ra.east |- si);

  \end{tikzpicture}
  \caption{
    The protocol designer specifies the protocol according to the protocol specification interface (Section~\ref{sec:spec:api}).
    Our tooling (the environment) derives an MDP and searches the optimal policy for selfish mining (Section~\ref{sec:mdp}).
  }
  \label{fig:spec_api}
\end{figure}
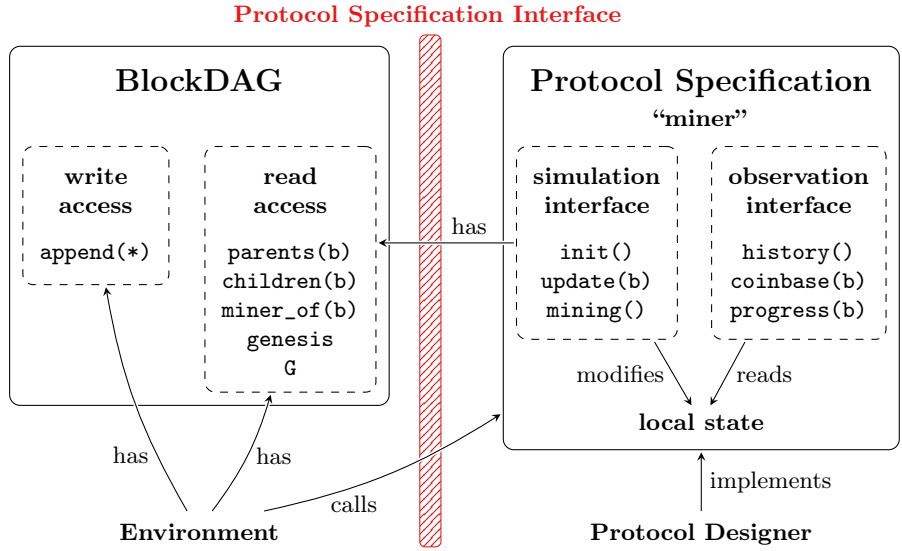

Our analysis framework is generic with respect to the protocol.
We here explain how the protocols are specified.
The other sections assume the protocol specification as an input.

The \emph{protocol designer} specifies the protocol by implementing the mining software in a constrained programming \emph{environment} which provides essential functionality like proof-of-work, reliable broadcast, and the BlockDAG.
The \emph{protocol implementation interface} is depicted in Figure~\ref{fig:spec_api}.

\subsection{Protocol Specification Interface}\label{sec:spec:api}

The protocol designer specifies the miner's behavior as Python functions, which the environment calls during execution.
We chose Python over pseudocode as this avoids ambiguity, enables automated testing and formatting, and closes the gap between the paper and the analytical code: the listings in this paper are the actual protocol implementations we feed into the automated analysis.

\paragraph{BlockDAG}\label{sec:spec:blockdag}

The specification can access the BlockDAG as follows.

\begin{description}
  \item[\texttt{parents(b)}, \texttt{children(b)}] return the parents and children of block \texttt{b}.
  \item[\texttt{miner\_of(b)}] returns the miner of block \texttt{b} and \texttt{me} identifies the miner themselves.
  \item[\texttt{genesis}] is the unique root block and \texttt{G} is the set of all blocks.
  \item[\texttt{height(b)}] returns the distance of block \texttt{b} to \texttt{genesis}.
\end{description}

\noindent Later, we use additional DAG-related terminology:
\begin{itemize}
  \item The \emph{past} and \emph{future} refer to the transitive closures of the parent and children relationships.
    The past of block $B$ includes the parents of $B$ as well as the parents of any other block in the past of $B$.
    The future is defined respectively for the children relationship.
  \item We say block $A$ confirms block $B$ to imply that $B$ is in the past of $A$.
\end{itemize}

Recall from Section~\ref{sec:attack_space}, that we restrict the visibility of blocks for the defender and let the attacker ignore blocks. This happens in a topological manner: \texttt{parents(b)} always returns all parents, \texttt{children(b)} and \texttt{G} might be subsets.

Also note that this interface does not allow direct modifications of the BlockDAG---only the environment can append new blocks.

\paragraph{Simulation Interface}

The protocol designer implements three functions: \texttt{init}, \texttt{update}, and \texttt{mining}, which together model the miner's behavior.
These functions can persist state across calls by setting attributes on the \texttt{state} object.
Attribute names can be chosen freely, but we assume the values are hashable.

\begin{description}
  \item[\texttt{init()}]
    initializes the miner's local \texttt{state}.
    The environment calls this at the beginning of the protocol execution.
    The function has no return value.
  \item[\texttt{update(b)}]
    updates the miner's local \texttt{state} with respect to a new block \texttt{b}.
    The environment calls this function to deliver blocks right after they become visible or stop being ignored.
    The function has no return value.
  \item[\texttt{mining()}]
    returns the parents of the next block to be mined. This specifies how the miner \emph{intends to} extend the BlockDAG.
    The actual extension is left to the environment and happens only if the miner was successful at proof-of-work.
\end{description}

Reflecting our assumptions in Section~\ref{sec:attack_space}, the protocol designer assumes that new blocks are automatically and reliably dissipated to all miners~(\ref{ass:comm}).
Delivery happens in a topological order~(\ref{ass:topo}).

\paragraph{Example (Bitcoin)}
In Bitcoin, the miner extends the longest chain of blocks, one block at a time. The \texttt{init()} function initializes \texttt{state.head = genesis}. The \texttt{update(b)} function checks whether the new block \texttt{b} has a longer history than \texttt{state.head}, and if so, updates \texttt{state.head = b}. The \texttt{mining()} function returns \texttt{state.head} as the single parent for the new block.
We specify this in Listing~\ref{alg:bitcoin}, Lines 1--9.

\begin{listing}
  \caption{Specification of the Bitcoin Protocol}
  \label{alg:bitcoin}
  \lstinput{bitcoin}
\end{listing}

\paragraph{Observation Interface}\label{par:spec:difficulty_adjustment}

The protocol designer implements three functions: \texttt{history}, \texttt{coinbase}, and \texttt{progress}, which inform the analyst about the miner's state. These functions can read but not modify the \texttt{state} object.

\begin{description}
  \item[\texttt{history()}]
    returns a list of blocks, the miner's opinion on the linear history.
  \item[\texttt{coinbase(b)}]
    defines the rewards allocated by block $b$. It returns a list of tuples \texttt{(r, v)}, where \texttt{r} denotes the recipient and \texttt{v} the size of the reward.
  \item[\texttt{progress(b)}]
    defines the feedback for difficulty adjustment.
    It returns a non-negative number defining the progress of block $b$.
    We assume the linear history grows at a constant rate, denoted in (expected) progress per time (\ref{ass:daa}).
\end{description}

\paragraph{Example (Bitcoin, cont'd)}
In Bitcoin, the linear history is the longest chain of blocks. For each block \texttt{b} in the linear history, \texttt{coinbase(b)} assigns a unit reward to the miner of \texttt{b}.
Similarly, there is one unit of progress per block.
We specify this in Listing~\ref{alg:bitcoin}, Lines 11--24.
Note, \texttt{set.pop()} removes and returns an arbitrary element of \texttt{set}.
This is deterministic because the set is singleton.

\subsection{Overview of the Specified Protocols}\label{sec:protocols}

Besides Bitcoin, we have specify Ethereum Proof-of-Work, its Byzantium update, GhostDAG~\cite{sompolinsky2021PhantomGhostdag}, and Parallel Proof-of-Work~\cite{keller2022ParallelProofofwork}.
We here provide short descriptions for each, shortly highlighting the core design ideas, and defer the full listings to Appendix~\ref{apx:protocols}.

\paragraph{Ethereum} can be seen as an extension to Bitcoin.
Like in Bitcoin, miners append their blocks to the longest chain, the history is the longest chain, and each block on the longest chain contributes one unit of progress for difficulty adjustment. New is that miners reference blocks off the longest chain as uncles and that these uncles get a full block reward. This compensates the miner's of blocks that would have missed their reward in Bitcoin.

Ethereum sets a time horizon $h$, expressed in numbers of blocks added to the longest chain, before which uncles have to be merged into the main chain.
Otherwise, they become orphans like in Bitcoin. The Ethereum cryptocurrency was deployed with $h = 7$, we set $h = 3$ throughout the paper.

We provide the full specification in Listing~\ref{alg:ethereum} in Appendix~\ref{apx:protocols}.

\paragraph{Byzantium} was deployed as an upgrade to the Ethereum cryptocurrency.
The overall chain structure stays intact, with the sole exception that a block can reference at most two uncles.
Besides that, Byzantium modifies the consensus rule, progress, and rewards.
Miners now extend the \emph{heaviest} chain, where uncles add the same weight as regular blocks.
Similarly, uncles add the same progress as regular blocks.
Uncle rewards are now discounted based on the distance to the block merging them into the chain, while the merging block gets a small bonus reward for each uncle.
Like Ethereum, Byzantium was deployed with horizon $h=7$.
We set horizon $h = 3$ throughout the paper.
The full specification is available in Listing~\ref{alg:byzantium} in Appendix~\ref{apx:protocols}.

\paragraph{GhostDAG} breaks with the notion of a longest or heaviest chain---miners just confirm all unconfirmed blocks.
The total order is established after the fact, for all blocks in the DAG.
GhostDAG comes with an heuristic for detecting longer phases of uncooperative behavior (private mining).
Blocks contributed by the (honest) majority are colored blue, the others red.
In the linear history, blue blocks take precedence over red blocks.
Blue blocks allocate one unit of reward to their miner, blue blocks none.
Similarly, blue blocks contribute one unit of progress to the chain, red blocks none.
Note that the publication~\cite{sompolinsky2021PhantomGhostdag} does not specify rewards and progress---we fill these gaps.

The heuristic for coloring blocks can be tuned through a natural number parameter $k$.
Higher $k$ allows for more parallelism in the DAG without punishing the minority miners.
The extreme setting $k=0$ roughly resembles Bitcoin.
We set $k=3$ throughout the paper.
The full specification is available in Listings~\ref{alg:ghostdag} and~\ref{alg:ghostdag_util} in Appendix~\ref{apx:protocols}.

\paragraph{Parallel Proof-of-Work}
uses two categories of blocks: votes and summaries.
Appending the next summary requires referencing $k$ votes confirming the parent summary.
The summaries form a linear chain and this chain constitutes the linear history.
As long as there are not enough votes for the next summary, miners append votes confirming the longest chain.
In case of ties, miners vote for the chain with the most votes already.
All blocks, both votes and summaries, assign a unit reward to their miner and contribute a unit progress to the difficulty adjustment.

Like in GhostDAG, higher $k$ allows for more parallelism.
The minimal setting $k = 1$ roughly resembles Bitcoin.
We set $k=3$ throughout the paper.

In the original proposal~\cite{keller2022ParallelProofofwork} summaries are not mined.
Instead there is a form of leader election: only the miner of the smallest vote (by hash) may append the next summary.
Later, the author has proposed a simplification~\cite{keller2023DAGStyleVoting} that works without public key cryptography.
Here, the summaries are mined like votes and this proof-of-work is the only form of authorization.
This is what we specify and analyse in this paper.
The full specification is available in Listing~\ref{alg:parallel} in Appendix~\ref{apx:protocols}.

\section{Automated Attack Search}\label{sec:mdp}

We now integrate the protocol-generic attack space from Section~\ref{sec:attack_space} with the protocol specification in Section~\ref{sec:spec}.
We will derive and solve MDPs that allow us to study the protocol's susceptibility to selfish mining, in a fully automated way.

Recall from Section~\ref{sec:background:mdp} that an MDP is a probabilistic state machine, where the inputs are called actions and the outputs are effects.
On each step of execution, the attacker observes the state and chooses the next action.
The attacker's objective is a function of the effects, in our case, maximizing rewards per time.

Assuming the protocol specification is given as input, the automated analysis pipeline is as follows.
We first define the transition function, which takes the current state and the chosen action as input and returns a discrete probability distribution for the next state (Section~\ref{sec:mdp:trans}).
After bounding the state space (Section~\ref{sec:mdp:finite}) and defining the transitions' effects (Section~\ref{sec:mdp:effect}), we derive an explicit tabular MDP (Section~\ref{sec:mdp:explore}).
Lastly, we define the objective function (Section~\ref{sec:mdp:objective}) and find the optimal policy (Section~\ref{sec:mdp:search}).

\subsection{State Transition Function}\label{sec:mdp:trans}

Recall from Section~\ref{sec:attack_space}, that we model two miners, attacker and defender.
The attacker ignores and withholds blocks using three types of actions \texttt{Consider(b)}, \texttt{Release(b)}, and \texttt{Continue}.
We describe the associated state transitions as a function that modifies the current state in response to the chosen action.

If the attacker chooses to \texttt{Consider(b)}, we remove block \texttt{b} from the set of ignored blocks and deliver it by calling the attacker's \texttt{update} function with argument \texttt{b}.
This transition is deterministic.
The action is only possible, if \texttt{b} is currently ignored and none of the parents of \texttt{b} are ignored.

\texttt{Release(b)} removes block \texttt{b} from the set of withheld blocks.
This is only possible, if \texttt{b} is currently withheld and none of the parents of \texttt{b} are withheld.
The transition is deterministic.
Note, the defender does not learn about \texttt{b} yet because communication is delayed.

\texttt{Continue} advances time until the next block is mined.
This action is always available.
It induces one out of four probabilistic transitions, depending on whether the attacker communicates quickly (probability $\gamma$) and who mines the next block (probability $\alpha$ for the attacker).

During the \texttt{Continue} action, the defender learns about all blocks that are neither withheld nor yet visible.
Depending on the attacker's luck regarding message reordering, we modify the order of delivery.
With probability $\gamma$, we deliver the attacker's blocks before the defender's blocks.
Otherwise, with probability $1-\gamma$, we deliver the defender's blocks first.
We deliver the blocks by calling the defender's \texttt{update} function.
This always happens in topological order~(\ref{ass:topo}).

After communication, we append a new block.
With probability $\alpha$, we call the attacker's \texttt{mining} function and mark the new block withheld and ignored.
Otherwise, with probability $1 - \alpha$, we call the defender's \texttt{mining} function and mark the new block ignored.
Note, even if the defender mines the new block, it only becomes visible during the next round of communication.
This enables us to model all honest system participants as a single miner.

As we lay out in Appendix~\ref{apx:traditional}, our communication assumptions align with the traditional selfish mining model against Bitcoin~\cite{sapirshtein2016OptimalSelfish}.
Similarly, the traditional actions, \emph{Wait}, \emph{Adopt}, \emph{Match}, and \emph{Override}, can be reproduced with a combination of our actions, \texttt{Consider(b)}, \texttt{Release(b)}, and \texttt{Continue}.

\subsection{State Space Cutoff}\label{sec:mdp:finite}

The state transition function induces an infinite number of states:
the BlockDAG grows monotonically, one block per \texttt{Continue} action.
Blocks are not removed and, throughout an episode, no state is visited twice.

Exact MDP solving techniques assume a finite number of states.
We thus impose a limit on the size of the BlockDAG (number of blocks).
In states that meet this limit, we force \emph{honest behavior}.
We make sure that all remaining policies loop within the restricted state space, by \emph{removing stale blocks} and \emph{truncating the common history}.
The details are as follows.

\paragraph*{Honest Behavior} We define the honest policy as follows:
\begin{enumerate}
  \item If any block $b$ is withheld, then \texttt{Release(b)}.
  \item If any block $b$ is ignored, then \texttt{Consider(b)}.
  \item Else, \texttt{Continue}.
\end{enumerate}
In states where the BlockDAG meets the size limit, we enforce this policy by taking away all other actions.
The attacker becomes honest and, by assumption~\ref{ass:consensus}, the miners will agree on the linear history.

\paragraph*{Truncating the Common History}

After applying the state transition function (Section~\ref{sec:mdp:trans}), we identify irrelevant parts of the BlockDAG and remove them.
Our heuristic is as follows:
\begin{enumerate}
  \item Obtain the linear histories of both miners, attacker and defender.
  \item Find the longest common prefix of the linear histories (common history).
  \item Find the last block \texttt{b} in the common history such that none of the blocks in \texttt{past(b)} have any children outside of $\{\texttt{b}\} \cup \texttt{past(b)}$.\label{enum:condition}
  \item Remove all blocks in \texttt{past(b)} from the BlockDAG.
\end{enumerate}

Point~\ref{enum:condition} preserves the semantics of the BlockDAG: block \texttt{b} becomes the new genesis (\ref{ass:genesis}) and none of the remaining blocks lose any of their parents.

Some protocols leave behind unconfirmed (stale) blocks, e.g., orphans in Bitcoin.
Stale blocks violate the condition in Point~\ref{enum:condition} and effectively prevent common history truncation.
We hence remove them in advance.

\paragraph*{Removing Stale Blocks}
We identify and remove state blocks as follows:

\begin{enumerate}
  \item \label{enum:stale:safeguard}
    Mark all blocks that are invisible to either miner, \texttt{H}onest or \texttt{A}ttacker.
  \item Mark all blocks that either miner would reference next (\texttt{mining} function).
  \item Mark all blocks that are in the past of any block marked before.
  \item Remove all \emph{unmarked} blocks from the BlockDAG.
\end{enumerate}

\noindent Point~\ref{enum:stale:safeguard} is a safeguard against removing blocks not relevant now but in the future.

\subsection{Effects: Reward and Progress}\label{sec:mdp:effect}

Each state transition has two effects: reward and progress.
We measure these on the common history, just before truncating it.
We iterate the blocks on the common history, apply the \texttt{coinbase} and \texttt{progress} functions to each, and accumulate the results.
The transition's reward is the accumulated reward assigned to the attacker.
The transition's progress is the accumulated progress.

\subsection{State Exploration}\label{sec:mdp:explore}

After state space truncation, the state transition function spans a finite attack space.
We now explore all reachable states, noting the available actions, and tabulating the probabilistic state transitions together with their effects.
This operation yields an \emph{explicit} (or tabular) MDP suitable for policy search.

Importantly, we evaluate the (expensive) state transition function once for each input.
Our attack search uses the cached results.
Also note, that we can update the model parameters, $\alpha$ and $\gamma$, in the explicit MDP without re-evaluating the state transition function, because only five distinct probabilities are involved.

\subsection{Objective Function: Reward per Progress}\label{sec:mdp:objective}

Selfish mining is about maximizing the expected reward per time, after the DAA has adjusted to the attack.
As discussed in Section~\ref{sec:attack_space}, the DAA creates a complex feedback loop between the miners' clocks, the overall hash rate, and the attacker's behavior.
Assumption~\ref{ass:daa} allows us break this loop: the DAA is in a stable state already and the expected progress per time fixed.
We hence can treat \emph{reward per progress} as a proxy for reward per time.

Note that Bitcoin, GhostDAG, and Parallel, assign exactly one unit of reward per unit of progress.
For these protocols, reward per progress equals relative reward, the objective function used in the traditional models~\cite{sapirshtein2016OptimalSelfish, barzur2020EfficientMDP}.

We use reward per progress to capture the differences to Ethereum and Byzantium.
In Ethereum, uncles are assigned full rewards but they do not count for progress.
Byzantium discounts uncle rewards, partially reallocating some of the discount to the merging block, while all blocks induce the same amount of progress.
For both protocols, relative reward does not capture all details.

\subsection{Probabilistic Termination Optimization}\label{sec:mdp:search}

Exact MDP solving algorithms expect that the objective is a linear function of the transition effects~\cite{sapirshtein2016OptimalSelfish, barzur2020EfficientMDP}.
Our objective function, reward per progress, is not linear.
However, as the same is true for relative reward, we can simply turn to the established solving techniques for the Bitcoin MDP~\cite{sapirshtein2016OptimalSelfish, barzur2020EfficientMDP}.

We use probabilistic termination optimization (PTO)~\cite{barzur2020EfficientMDP}.
In short, the trick is to transform the (source) MDP, where all policies loop, into an MDP where all episodes terminate.
In the terminating MDP, the expected progress per episode (horizon) is sufficiently independent from the policy.
The authors prove that, as the horizon grows to infinity, any reward-maximizing policy in the terminating MDP also maximizes the reward per progress in the source MDP.

We derive terminating MDPs with horizon 100.
Then we apply a standard dynamic programming algorithm, value iteration, to obtain an $10^{-4}$-optimal policy.
We then apply this policy to the \emph{source} MDP and find its steady state using linear equations.
After calculating the expected reward \emph{and} progress at the steady state, we obtain the expected reward \emph{per} progress by dividing those numbers.
This is what we report in Section~\ref{sec:results}.

\section{State Space Optimizations}\label{sec:optimizations}

We propose two optimizations to reduce the size of the state space: canonizing the BlockDAG and forcing the attacker to consider their own blocks.

\paragraph*{Canonization}

We use adjacency lists to track the evolution of the BlockDAG over time.
The blocks are enumerated in the order in which they were mined (compare Figure~\ref{fig:state}).
However, for proof-of-work consensus protocols, the exact mining order does not matter.
In fact, the true mining order is unknown to the participants;
they rely on consensus to establish an order in the first place.

So, our default BlockDAG implementation is bound to create redundant states.
E.g., the blocks in Figure~\ref{fig:state} can be relabelled in 69 ways without breaking the topological constraints (parents have lower ids than their children) and there are 35 options for the blocks in Figure~\ref{fig:state_btc} in Appendix~\ref{apx:traditional}.

We optionally merge these states during state exploration (Section~\ref{sec:mdp:explore}) by relabelling the blocks canonically:
We first derive a vertex coloring from the relevant block properties (who mined it and whether it is visible/ignored/withheld).
We then find a canonical and color-preserving vertex labeling using appropriate third-party software, Nauty~\cite{mckay2014PracticalGraph}.
The protocol designer may provide assistive input for canonization by implementing a function that colors blocks depending on the miner's local state.

Note that the GhostDAG protocol assumes a deterministic topological ordering of blocks.
We specify this using Python's builtin \texttt{hash} function (Lines 13 and 18 of Listing~\ref{alg:ghostdag} in Appendix~\ref{apx:protocols}).
Canonization does not preserve these hashes (blocks are integers) and thus is an invalid operation for GhostDAG.

\paragraph*{Force Consider}

The traditional selfish mining model against Bitcoin~\cite{sapirshtein2016OptimalSelfish} allows at most two distinct chains (one fork).
In our attack model, the attacker can create many forks, by temporally ignoring their own blocks.

We optionally adopt the traditional behavior and force the attacker to consider their own blocks.
This happens during the same transition in which the new block has been mined (as opposed to enforcing a consider action after the transition).

\section{Results}\label{sec:results}

We now apply the automatic analysis pipeline (Section~\ref{sec:mdp}) to the specified protocols (Section~\ref{sec:protocols}).
We first focus on the size of the state space (Section~\ref{sec:res:sss}).
Then we validate our modeling choices (Section~\ref{sec:attack_space} and~\ref{sec:mdp}) against established models for Bitcoin (Section~\ref{sec:res:validation}).
This is followed by a comparison of the protocols to determine which is the least susceptible to selfish mining (Section~\ref{sec:res:protos}).
Lastly, we evaluate whether and how our state space optimizations (Section~\ref{sec:optimizations}) affect the results on selfish mining~(Section~\ref{sec:res:optim}).

\paragraph*{Reference Models}

Throughout this section, we compare our generic model to the traditional models specific to Bitcoin~\cite{sapirshtein2016OptimalSelfish, barzur2020EfficientMDP}.
We obtain the reported numbers from implementing the state transition functions and effects as described in the respective papers (comp.~Section~\ref{sec:mdp:trans} and~\ref{sec:mdp:effect}).
The remaining parts of the analysis are the same across models: we restrict the state space based on the number of blocks (Section~\ref{sec:mdp:finite}), build a tabular MDP (Section~\ref{sec:mdp:explore}), and optimize the policy for selfish mining (Sections~\ref{sec:mdp:objective} and~\ref{sec:mdp:search}).

\subsection{Size of the State Space}\label{sec:res:sss}

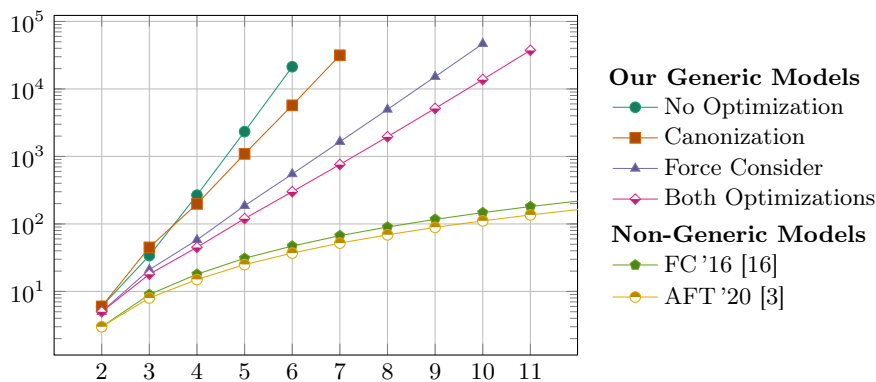
\begin{figure}[tbp]
  \centering
  \def\data{fig/state-space-size-btc.csv}
  \begin{tikzpicture}
    \begin{axis}[
      width=0.7\textwidth,
      height=0.5\textwidth,
      xtick=data,
      xmax=11.99, %
      ymode=log,
      legend cell align={left},
      legend style={at={(1.05,0.5)}, anchor=west, legend columns=1, style={draw=none}},
      grid=major,
      cycle list/Dark2, %
      cycle multiindex* list={
        Dark2\nextlist
        mark list* \nextlist %
      },
      ]

      \addplot[draw=none, forget plot] table [x=dag_size_limit, y=ref-fc16, col sep=comma, draw=none] {\data};

      \addlegendimage{empty legend}
      \addlegendentry[yshift=5pt]{\strut}
      \addlegendimage{empty legend}
      \addlegendentry[xshift=-21pt, yshift=5pt, overlay]{\textbf{Our Generic Models}}

      \addplot table [x=dag_size_limit, y=v1, col sep=comma] {\data};
      \label{plot:sss_btc:baseline}
      \addlegendentry{No Optimization}

      \addplot table [x=dag_size_limit, y=v1+n, col sep=comma] {\data};
      \label{plot:sss_btc:canon}
      \addlegendentry{Canonization}

      \addplot table [x=dag_size_limit, y=v1+fc, col sep=comma] {\data};
      \label{plot:sss_btc:consider}
      \addlegendentry{Force Consider}

      \addplot table [x=dag_size_limit, y=v1+fc+n, col sep=comma] {\data};
      \label{plot:sss_btc:canon_consider}
      \addlegendentry{Both Optimizations}

      \addlegendimage{empty legend}
      \addlegendentry[yshift=5pt]{\strut}
      \addlegendimage{empty legend}
      \addlegendentry[xshift=-20pt, yshift=5pt, overlay]{\textbf{Non-Generic Models}}

      \addplot table [x=dag_size_limit, y=ref-fc16, col sep=comma] {\data};
      \label{plot:sss_btc:reference}
      \addlegendentry{FC\,'16\cite{sapirshtein2016OptimalSelfish}}

      \addplot table [x=dag_size_limit, y=ref-aft20, col sep=comma] {\data};
      \label{plot:sss_btc:reference}
      \addlegendentry{AFT\,'20\cite{barzur2020EfficientMDP}}
    \end{axis}
  \end{tikzpicture}
  \caption{
    Number of states of the Bitcoin MDP ($y$-axis) as a function of the BlockDAG's size limit ($x$-axis, number of blocks).
    We compare our model with optimizations turned on and off and add the traditional non-generic models~\cite{sapirshtein2016OptimalSelfish, barzur2020EfficientMDP} as a reference.
  }
  \label{fig:sss_btc}
\end{figure}

Recall from Section~\ref{sec:mdp:finite}, that we limit the number of blocks in the BlockDAG to bound the number of states in the MDP.
We expect the state size to grow as we increase the BlockDAG size limit.
We now evaluate how fast.

We do this, by applying the analysis pipeline to the specified protocols, skipping the attack search itself.
Instead we observe the number of states in the resulting MDP.
We start with a BlockDAG size limit of 2 and increase it one-by-one, until the state size exceeds 100\,000 states.
This upper limit ensures that we can reproduce all results, including the following sections, within five hours on our consumer grade laptop ($\approx\$\,1\,000$).

Figure~\ref{fig:sss_btc} visualizes the growth of the state space for Bitcoin, with state space optimizations turned on and off (Section~\ref{sec:optimizations}).
The optimizations are effective: when applying them both, we exceed the state size limit after 11 instead of 6 blocks.
As a reference, we compare our generic model to the traditional models specific to Bitcoin~\cite{sapirshtein2016OptimalSelfish, barzur2020EfficientMDP}.
Expectedly, the specific models have much smaller state sizes.
Note that the plot is truncated on the right: the traditional models exceed the state size limit after 239 and 283 blocks respectively.

\begin{table}[b]
  \def\n{\phantom{0}}
  \newcolumntype{Y}{{\raggedright\arraybackslash}X}
  \caption{%
    State size scaling for all protocols and state space optimizations.
  }
  \label{tab:state_size}
  \begin{tabularx}{\linewidth}{X cc cc cc cc}
    \toprule
    & \multicolumn{2}{c}{No Opt.}
    & \multicolumn{2}{c}{Canonization}
    & \multicolumn{2}{c}{Force Consider}
    & \multicolumn{2}{c}{Both Opt.} \\
    \cmidrule(lr){2-3}
    \cmidrule(lr){4-5}
    \cmidrule(lr){6-7}
    \cmidrule(lr){8-9}
    Protocol &
    $s_{\operatorname{max}}$ & $n_6$ &
    $s_{\operatorname{max}}$ & $n_6$ &
    $s_{\operatorname{max}}$ & $n_6$ &
    $s_{\operatorname{max}}$ & $n_6$ \\
    \midrule
    Bitcoin   & 6 &  21327 &  7 & \n5724 &  10 & \n549 &  11 &  300 \\
    Ethereum  & 6 &  32961 &  7 &  10293 & \n9 &  1179 &  10 &  596 \\
    Byzantium & 6 &  33016 &  7 & \n9879 & \n9 &  1109 &  10 &  572 \\
    GhostDAG  & 6 &  46966 &    &        & \n8 &  1527 &     &      \\
    Parallel  & 7 & \n9122 &  8 & \n1654 & \n8 &  2050 &  12 &  462 \\
    \bottomrule
  \end{tabularx}

  \smallskip
  We grow the BlockDAG size limit (number of blocks), until the resulting model exceeds 100\,000 states.
  We here report the maximum BlockDAG size limit ($s_{\operatorname{max}}$) before reaching this limit, as well as the number of states ($n_6$) at a fixed size limit of 6 blocks.
  We omit the results for the invalid optimization, canonization on GhostDAG.
\end{table}

Table~\ref{tab:state_size} expands this analysis for the other protocols, excluding the invalid optimization, canonization on GhostDAG. %
We report the maximum number of blocks explored ($s_{\operatorname{max}}$) before exceeding the state size limit of 100\,000 states, and the number of states ($n_6$) when limiting the BlockDAG size to 6 blocks.
For reference, we could explore the traditional model~\cite{sapirshtein2016OptimalSelfish} up to $s_{\operatorname{max}} = 239$ before reaching 100\,000 states, and it has $n_6 = 47$ states when limiting the BlockDAG to 6 blocks.
For the second reference model~\cite{barzur2020EfficientMDP}, $s_{\operatorname{max}} = 283$ and $n_6 = 37$.

\subsection{Validation Against Traditional Models for Bitcoin}\label{sec:validation}\label{sec:res:validation}

\begin{figure}[tb]
  \raggedleft
  \begin{tikzpicture}
    \begin{groupplot}[
      group style={
        group size=2 by 1,
        ylabels at=edge left,
        yticklabels at=edge left,
        horizontal sep=1ex,
      },
      width=0.59\textwidth,
      height=0.4\textwidth,
      xtick=data,
      ytick={0.33, 0.35, 0.37, 0.39, 0.41, 0.43},
      legend cell align={left},
      legend style={at={(1.05,0.5)}, anchor=west, legend columns=1, style={draw=none}},
      grid=major,
      cycle list/Dark2, %
      cycle multiindex* list={
        Dark2\nextlist
        mark list* \nextlist %
      },
      ]

      \def\alignyaxes{
        \def\data{fig/rpp-for-dag-size-btc-alpha33-gamma66.csv}
        \addplot[draw=none, forget plot] table [x=dag_size_limit, y=ref-fc16, col sep=comma] {\data};
      }

      \nextgroupplot[title={$\gamma = 0.33$}]
      \alignyaxes

      \def\data{fig/rpp-for-dag-size-btc-alpha33-gamma33.csv}

      \addplot table [x=dag_size_limit, y=v1+fc+n, col sep=comma] {\data};
      \label{plot:rpp_btc:our}

      \addplot table [x=dag_size_limit, y=ref-fc16, col sep=comma] {\data};
      \label{plot:rpp_btc:fc16}

      \addplot table [x=dag_size_limit, y=ref-aft20, col sep=comma] {\data};
      \label{plot:rpp_btc:aft20}

      \nextgroupplot[title={$\gamma = 0.66$}]
      \alignyaxes

      \def\data{fig/rpp-for-dag-size-btc-alpha33-gamma66.csv}

      \addplot table [x=dag_size_limit, y=v1+fc+n, col sep=comma] {\data};

      \addplot table [x=dag_size_limit, y=ref-fc16, col sep=comma] {\data};

      \addplot table [x=dag_size_limit, y=ref-aft20, col sep=comma] {\data};

    \end{groupplot}
  \end{tikzpicture}
  \caption{
    Reward per progress of the optimal selfish mining policy against Bitcoin ($y$-axis) as a function of the BlockDAG size limit ($x$-axis, number of blocks).
    We compare our model (\ref{plot:rpp_btc:our}) to the non-generic models presented at FC\,'16 \cite{sapirshtein2016OptimalSelfish} (\ref{plot:rpp_btc:fc16}) and AFT\,'20 \cite{barzur2020EfficientMDP} (\ref{plot:rpp_btc:aft20}).
    We set the attacker's relative hash rate $\alpha = 0.33$ and communication advantage $\gamma \in \{0.33, 0.66\}$ (left/right).
    On our model, we enable BlockDAG canonization and force the attacker to consider their own blocks.
    Line $y = 0.33$ represents honest behavior.
  }
  \label{fig:rpp_btc}
\end{figure}
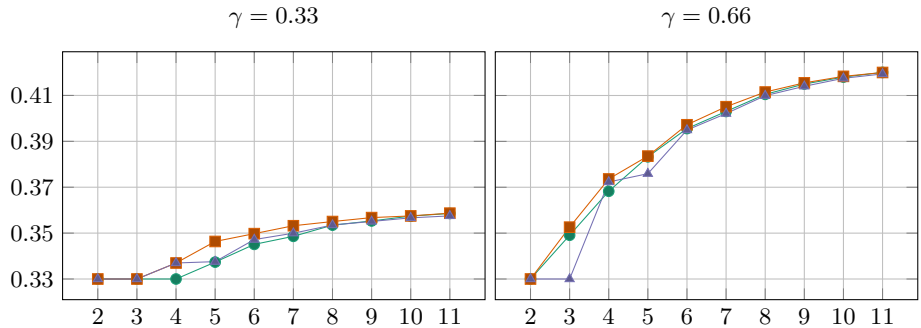

We now validate our protocol-generic model against existing models specific to Bitcoin~\cite{sapirshtein2016OptimalSelfish, barzur2020EfficientMDP}.
We take 30 MDPs from Figure~\ref{fig:sss_btc}: our model with both optimizations turned on, the two reference models, and BlockDAG size limits ranging from 2 to 11 blocks.
We then apply the remaining analysis, policy optimization and evaluation (Section~\ref{sec:mdp:search}), with model parameters $\alpha = 0.33$ and $\gamma \in \{0.33, 0.66\}$.
Figure~\ref{fig:rpp_btc} reports the results, denoted in (expected) reward per progress.

We make three observations:
First, reward per progress grows monotonically for increasing BlockDAG size limit, independent of the model.
This indicates that the BlockDAG size limit restricts the attacker.
In a real setting, there is no BlockDAG size limit and the attacker's reward per progress could be even higher.
Second, using the FC\,'16 model~\cite{sapirshtein2016OptimalSelfish} as baseline, the error of our generic model is comparably sized to the error of the AFT\,'20 model~\cite{barzur2020EfficientMDP}.
Third, the effect of increasing the BlockDAG size limit by one, e.g. from 10 to 11, dominates the remaining differences between the models.
All three models appear to converge on the same result.

\subsection{Comparison of Protocols}\label{sec:res:protos}

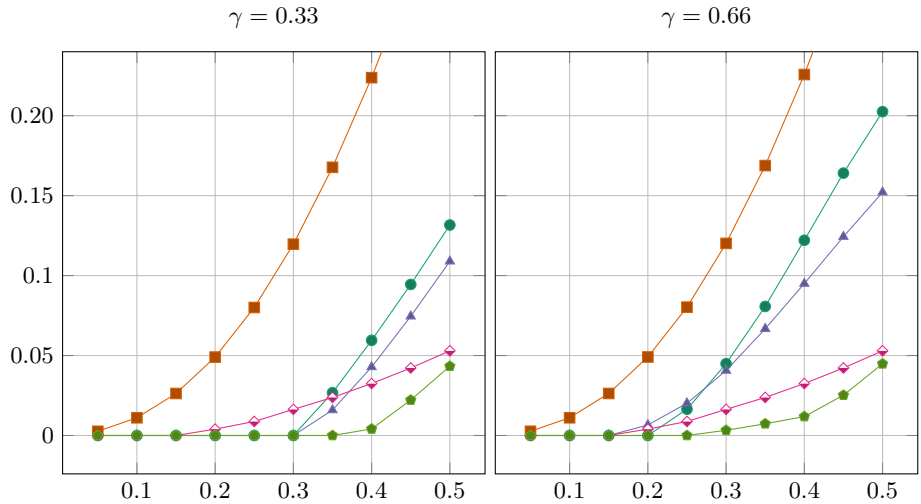
\begin{figure}[tb]
  \raggedleft
  \def\data{fig/rpp-of-alpha.csv}
  \begin{tikzpicture}
    \begin{groupplot}[
      group style={
        group size=2 by 1,
        xlabels at=edge bottom,
        xticklabels at=edge bottom,
        ylabels at=edge left,
        yticklabels at=edge left,
        horizontal sep=1ex,
      },
      width=0.59\textwidth,
      height=0.59\textwidth,
      ytick={0, 0.05, 0.1, 0.15, 0.20},
      yticklabels={0, 0.05, 0.1, 0.15, 0.20},
      ymax=0.24,
      legend cell align={left},
      legend style={at={(1.05,0.5)}, anchor=west, legend columns=1, style={draw=none}},
      grid=major,
      cycle list/Dark2, %
      cycle multiindex* list={
        Dark2\nextlist
        mark list* \nextlist %
      },
      ]

      \def\alignyaxes{
      }

      \nextgroupplot[title={$\gamma = 0.33$}]
      \alignyaxes

      \addplot table [x=alpha, y expr=\thisrow{v1:bitcoin:gamma33:dsl6} - \thisrow{alpha}, col sep=comma] {\data};
      \label{plot:rpp_alpha:bitcoin}
      \addplot table [x=alpha, y expr=\thisrow{v1:ethereum_3:gamma33:dsl6} - \thisrow{alpha}, col sep=comma] {\data};
      \label{plot:rpp_alpha:ethereum}
      \addplot table [x=alpha, y expr=\thisrow{v1:byzantium_3:gamma33:dsl6} - \thisrow{alpha}, col sep=comma] {\data};
      \label{plot:rpp_alpha:byzantium}
      \addplot table [x=alpha, y expr=\thisrow{v1:ghostdag_3:gamma33:dsl6} - \thisrow{alpha}, col sep=comma] {\data};
      \label{plot:rpp_alpha:ghostdag}
      \addplot table [x=alpha, y expr=\thisrow{v1:parallel_3:gamma33:dsl6} - \thisrow{alpha}, col sep=comma] {\data};
      \label{plot:rpp_alpha:parallel}

      \nextgroupplot[title={$\gamma = 0.66$}]
      \alignyaxes

      \addplot table [x=alpha, y expr=\thisrow{v1:bitcoin:gamma66:dsl6} - \thisrow{alpha}, col sep=comma] {\data};
      \label{plot:rpp_alpha:bitcoin}
      \addplot table [x=alpha, y expr=\thisrow{v1:ethereum_3:gamma66:dsl6} - \thisrow{alpha}, col sep=comma] {\data};
      \label{plot:rpp_alpha:ethereum}
      \addplot table [x=alpha, y expr=\thisrow{v1:byzantium_3:gamma66:dsl6} - \thisrow{alpha}, col sep=comma] {\data};
      \label{plot:rpp_alpha:byzantium}
      \addplot table [x=alpha, y expr=\thisrow{v1:ghostdag_3:gamma66:dsl6} - \thisrow{alpha}, col sep=comma] {\data};
      \label{plot:rpp_alpha:ghostdag}
      \addplot table [x=alpha, y expr=\thisrow{v1:parallel_3:gamma66:dsl6} - \thisrow{alpha}, col sep=comma] {\data};
      \label{plot:rpp_alpha:parallel}

    \end{groupplot}
  \end{tikzpicture}
  \caption{
    Surplus of the optimal policy ($y$-axis, reward per progress minus $\alpha$) as a function of the attacker's relative hash rate $\alpha$ ($x$-axis).
    We compare the protocols Bitcoin~(\ref{plot:rpp_alpha:bitcoin}), Ethereum~(\ref{plot:rpp_alpha:ethereum}), Byzantium~(\ref{plot:rpp_alpha:byzantium}), GhostDAG~(\ref{plot:rpp_alpha:ghostdag}), and Parallel Proof-of-Work~(\ref{plot:rpp_alpha:parallel}).
    All results are obtained from our generic attack model with optimizations turned off.
    We set communication advantage $\gamma \in \{0.33, 0.66\}$ (left/right) and limit the BlockDAG to at most 6 blocks.
    Line $ y = 0 $ represents honest behavior.
  }
  \label{fig:rpp_all}
\end{figure}

We now evaluate selfish mining across all specified protocols: Bitcoin, Ethereum, Byzantium, GhostDAG, and Parallel Proof-of-Work.
We take the 5 MDPs from Table~\ref{tab:state_size}, where the BlockDAG is limited to 6 blocks ($n_6$ column) and both optimizations are turned off.
We then apply policy optimization and evaluation (Section~\ref{sec:mdp:search}), with model parameters $\alpha \in \{0.05, 0.1, \dots, 0.45, 0.5\}$ and $\gamma \in \{0.33, 0.66\}$.
Figure~\ref{fig:rpp_all} reports the results, denoted in (expected) reward per progress minus $\alpha$.
This represents the \emph{surplus} of selfish mining as compared to honest behavior.
Note that we truncate the $\alpha = 0.5$ result for Ethereum, to give more detail to the other protocols.
The truncated result can be read from Figure~\ref{fig:rpp_fc} of the next experiment.

This comparison creates important insights for protocol design.
First, the first Ethereum protocol is clearly off the table.
Second, the fixes deployed with the Byzantium upgrade are effective.
Compared to Bitcoin, Byzantium is less susceptible to selfish mining for most combinations of $\alpha$ and $\gamma$, and the difference is relatively small if not.
Third, GhostDAG seems to be incentive incompatible for relatively weak attackers (low $\alpha$) although it clearly outperforms Bitcoin and Byzantium for strong attackers.
Forth, Parallel Proof-of-Work seems to hit a sweet spot, showing the lowest reward per progress for all parameter combinations.

This warrants further research.
We are particularly interested in how GhostDAG's and Parallel's $k$-parameters affect the results and to which extent the results are skewed by our low BlockDAG size limit.
We here set $k=3$, while the designers of both protocols propose much higher values.

\subsection{State Space Optimizations}\label{sec:res:optim}

We have presented two attack space optimizations in Section~\ref{sec:optimizations}: canonization of the BlockDAG and forcing the attacker to consider their own blocks.
We now check whether these optimizations affect the reward per progress after policy optimization.
We take the 18 MDPs of Table~\ref{tab:state_size}, where the BlockDAG is limited to 6 blocks (all four $n_6$ columns).
We then apply policy optimization and evaluation (Section~\ref{sec:mdp:search}), with model parameters $\alpha \in \{0.05, 0.1, \dots, 0.45, 0.5\}$ and $\gamma \in \{0.33, 0.66\}$.

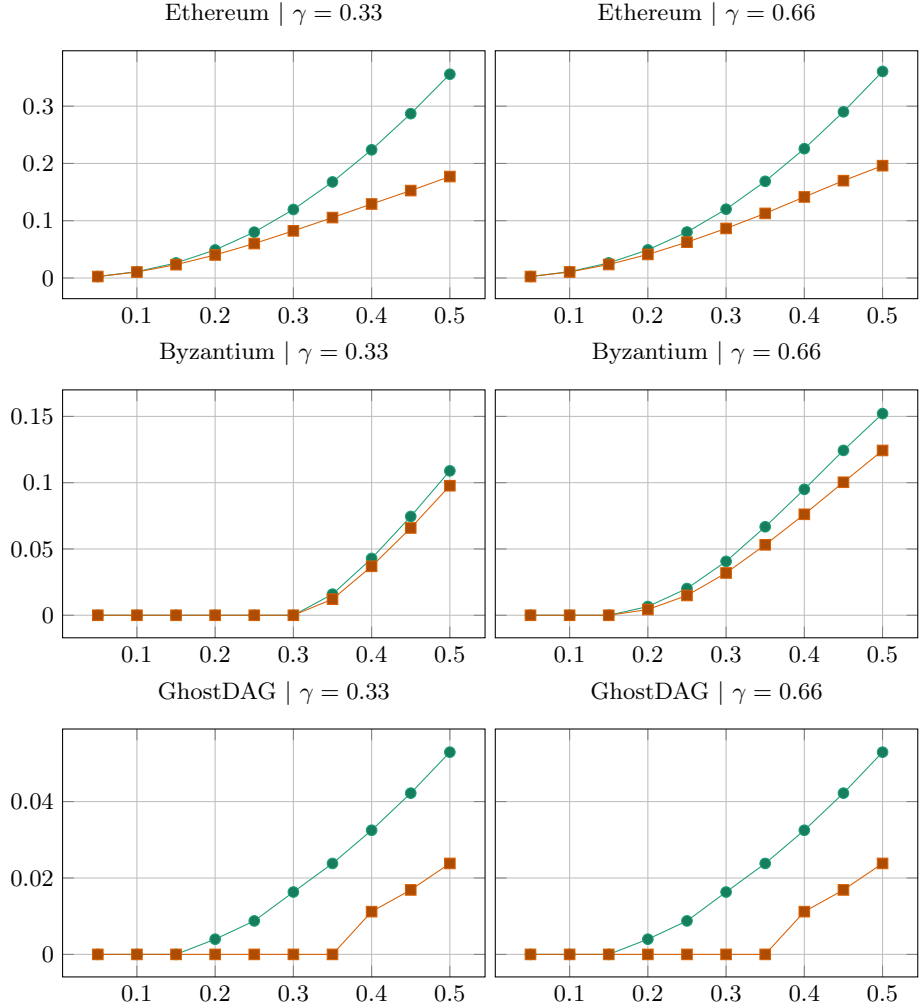
\begin{figure}[tbp]
  \raggedleft
  \def\data{fig/rpp-of-alpha.csv}
  \begin{tikzpicture}
    \begin{groupplot}[
      group style={
        group size=2 by 1,
        xlabels at=edge bottom,
        xticklabels at=edge bottom,
        ylabels at=edge left,
        yticklabels at=edge left,
        horizontal sep=1ex,
      },
      width=0.59\textwidth,
      height=0.4\textwidth,
      legend cell align={left},
      legend style={at={(1.05,0.5)}, anchor=west, legend columns=1, style={draw=none}},
      grid=major,
      cycle list/Dark2, %
      cycle multiindex* list={
        Dark2\nextlist
        mark list* \nextlist %
      },
      ]

      \def\alignyaxes{
        \addplot[draw=none, forget plot] table [x=alpha, y expr = \thisrow{v1:ethereum_3:gamma33:dsl6} - \thisrow{alpha}, col sep=comma] {\data};
        \addplot[draw=none, forget plot] table [x=alpha, y expr = \thisrow{v1:ethereum_3:gamma66:dsl6} - \thisrow{alpha}, col sep=comma] {\data};
        \addplot[draw=none, forget plot] table [x=alpha, y expr = \thisrow{v1+fc:byzantium_3:gamma33:dsl6} - \thisrow{alpha}, col sep=comma] {\data};
      }

      \nextgroupplot[title={Ethereum $|$ $\gamma = 0.33$}]
      \alignyaxes

      \addplot table [x=alpha, y expr = \thisrow{v1:ethereum_3:gamma33:dsl6} - \thisrow{alpha}, col sep=comma] {\data};
      \label{plot:eth:rpp_alpha:fcoff}
      \addplot table [x=alpha, y expr = \thisrow{v1+fc:ethereum_3:gamma33:dsl6} - \thisrow{alpha}, col sep=comma] {\data};
      \label{plot:eth:rpp_alpha:fcon}

      \nextgroupplot[title={Ethereum $|$ $\gamma = 0.66$}]
      \alignyaxes

      \addplot table [x=alpha, y expr = \thisrow{v1:ethereum_3:gamma66:dsl6} - \thisrow{alpha}, col sep=comma] {\data};
      \addplot table [x=alpha, y expr = \thisrow{v1+fc:ethereum_3:gamma66:dsl6} - \thisrow{alpha}, col sep=comma] {\data};

    \end{groupplot}
  \end{tikzpicture}
  \begin{tikzpicture}
    \begin{groupplot}[
      group style={
        group size=2 by 1,
        xlabels at=edge bottom,
        xticklabels at=edge bottom,
        ylabels at=edge left,
        yticklabels at=edge left,
        horizontal sep=1ex,
      },
      width=0.59\textwidth,
      height=0.4\textwidth,
      ytick={0, 0.05, 0.1, 0.15},
      yticklabels={0, 0.05, 0.1, 0.15},
      ymax=0.17,
      legend cell align={left},
      legend style={at={(1.05,0.5)}, anchor=west, legend columns=1, style={draw=none}},
      grid=major,
      cycle list/Dark2, %
      cycle multiindex* list={
        Dark2\nextlist
        mark list* \nextlist %
      },
      ]

      \nextgroupplot[title={Byzantium $|$ $\gamma = 0.33$}]

      \addplot table [x=alpha, y expr = \thisrow{v1:byzantium_3:gamma33:dsl6} - \thisrow{alpha}, col sep=comma] {\data};
      \addplot table [x=alpha, y expr = \thisrow{v1+fc:byzantium_3:gamma33:dsl6} - \thisrow{alpha}, col sep=comma] {\data};

      \nextgroupplot[title={Byzantium $|$ $\gamma = 0.66$}]

      \addplot table [x=alpha, y expr = \thisrow{v1:byzantium_3:gamma66:dsl6} - \thisrow{alpha}, col sep=comma] {\data};
      \addplot table [x=alpha, y expr = \thisrow{v1+fc:byzantium_3:gamma66:dsl6} - \thisrow{alpha}, col sep=comma] {\data};

    \end{groupplot}
  \end{tikzpicture}
  \begin{tikzpicture}
    \begin{groupplot}[
      group style={
        group size=2 by 1,
        xlabels at=edge bottom,
        xticklabels at=edge bottom,
        ylabels at=edge left,
        yticklabels at=edge left,
        horizontal sep=1ex,
      },
      width=0.59\textwidth,
      height=0.4\textwidth,
      ytick={0, 0.02, 0.04},
      yticklabels={0, 0.02, 0.04},
      ymax=0.059,
      scaled y ticks = false,
      legend cell align={left},
      legend style={at={(1.05,0.5)}, anchor=west, legend columns=1, style={draw=none}},
      grid=major,
      cycle list/Dark2, %
      cycle multiindex* list={
        Dark2\nextlist
        mark list* \nextlist %
      },
      ]

      \nextgroupplot[title={GhostDAG $|$ $\gamma = 0.33$}]

      \addplot table [x=alpha, y expr = \thisrow{v1:ghostdag_3:gamma33:dsl6} - \thisrow{alpha}, col sep=comma] {\data};
      \addplot table [x=alpha, y expr = \thisrow{v1+fc:ghostdag_3:gamma33:dsl6} - \thisrow{alpha}, col sep=comma] {\data};

      \nextgroupplot[title={GhostDAG $|$ $\gamma = 0.66$}]

      \addplot table [x=alpha, y expr = \thisrow{v1:ghostdag_3:gamma66:dsl6} - \thisrow{alpha}, col sep=comma] {\data};
      \addplot table [x=alpha, y expr = \thisrow{v1+fc:ghostdag_3:gamma66:dsl6} - \thisrow{alpha}, col sep=comma] {\data};

    \end{groupplot}
  \end{tikzpicture}
  \caption{
    Surplus of the optimal policy ($y$-axis, reward per progress minus $\alpha$) as a function of the attacker's relative hash rate $\alpha$ ($x$-axis).
    We compare our generic model \textbf{with} (\ref{plot:eth:rpp_alpha:fcon}) and \textbf{without} (\ref{plot:eth:rpp_alpha:fcoff}) \textbf{forcing the attacker to consider their own blocks}.
    We report the results for Ethereum (top), its Byzantium upgrade (middle), and GhostDAG (bottom); the other protocols, Bitcoin and Parallel Proof-of-Work, are not sensitive to this optimization.
    Note the different scales on the $y$-axis and see Figure~\ref{fig:rpp_all} for comparing results across protocols.
    We set communication advantage $\gamma \in \{0.33, 0.66\}$ (left/right) and limit the BlockDAG to at most 6 blocks.
    Line $y = 0$ represents honest behavior.
  }
  \label{fig:rpp_fc}
\end{figure}

We do not observe substantial differences with the \emph{Canonization} optimization turned on or off.
Some protocols, however, are sensitive to the \emph{Force Consider} optimization: Ethereum, Byzantium, and GhostDAG.
Figure~\ref{fig:rpp_fc} shows the differences for these protocols with \emph{Force Consider} turned on and off.
As before, we report the (expected) reward per progress minus $\alpha$, which represents the \emph{surplus} of selfish mining as compared to honest behavior.

We observe that \emph{Force Consider} reduces the surplus.
This suggest that the optimal selfish mining strategy against Ethereum, Byzantium, and GhostDAG involves ignoring own blocks, at least temporally.
We also conclude that the \emph{Force Consider} optimization has to be applied with care.

\section{Discussion and Conclusion}\label{sec:discussion}

Our implementation is optimized for correctness and ease of presentation.
The results reproduce on a \$\,1\,000 laptop in a few hours.
This comes at the cost of restricting the problem size (at most 100\,000 states) and the size of the BlockDAG (6--11 blocks depending on model and protocol).
This restricts the applicability of our tooling.
E.g., the configuration of Parallel Proof-of-Work proposed by its authors~\cite{keller2022ParallelProofofwork} requests $k=51$ parallel votes before growing the linear history by a single block.
Restricting the size of the BlockDAG to anything below $k$ is futile, and hence the proposed configuration escapes our treatment.

This raises the question, how far the problem size can be scaled after optimizing the implementation and reaching for more capable hardware.
But considering the exponential growth in Figure~\ref{fig:sss_btc}, we think algorithmic advances are inevitable.
Modern reinforcement learning algorithms, e.g. DQN or PPO, should be an option, as they have been applied to similar problems before~\cite{barzur2022WeRLmanTackle, keller2023TailstormSecure}.
In principle, such techniques allow for unbounded state spaces, potentially opening a path to removing the BlockDAG size limit altogether.
Beware though, that this comes at the cost of optimality.
Our approach provides optimal policies and can serve as a baseline for evaluating approximate results in the future.

This brings us to other avenues for future research.
We here measure two effects of the attacker's policy, reward and progress on the common chain (Section~\ref{sec:mdp:effect}), and analyse a single objective function, reward per progress (Section~\ref{sec:mdp:objective}), modeling the long term incentives after the DAA has adjusted to the policy.
We deem it worthwhile to observe more effects, e.g., number of blocks mined and number of blocks rewritten on the defender's chain, to enable other relevant objective functions, e.g., for modelling consistency, liveness, and short term incentives.

We also see low hanging fruits to extend the range of protocols supported.
E.g., some protocols use a ``two-for-one trick'' where the proof-of-work puzzle can have multiple possible solutions~\cite{pass2017FruitChainsFair, abraham2023ColordagIncentivecompatible}.
This can be modeled by coloring the blocks probabilistically, just when they are appended to the BlockDAG.
The same mechanism would allow to (approximately) model hash-based precedence rules~\cite{sompolinsky2021PhantomGhostdag, keller2023TailstormSecure} which, we think, can be gamed.
Turning to precedence rules, the random-tie breaking proposed by the inventors of selfish mining themselves~\cite{eyal2014MajorityNot}, to reduce the negative effects of high communication advantages $\gamma$ on Bitcoin, should be an interesting case study as well.
Specifying this would require probabilistic state \texttt{update} functions---simple as long as the number of outcomes is finite.

\paragraph*{Conclusion}
We have presented a framework for finding optimal selfish mining attacks against a wide range of proof-of-work consensus protocols.
Our work reproduces established results for well studied protocols, Bitcoin and Ethereum, and also covers protocols that have not been analyzed before (GhostDAG) or only with approximate policy optimization techniques (Parallel Proof-of-Work).
A key feature of our approach is its modularity: the protocol designer specifies the protocol as a concise Python program, the subsequent analysis, including generating the selfish mining MDP and optimizing the policy, follows automatically.
This modularity, combined with the release of all our tooling~\cite{code}, enables rapid iteration during protocol design.

\section*{Acknowledgements}
This work is inspired by a previous collaboration with George Bissias~\cite{keller2023TailstormSecure}.
I would like to thank Rainer Böhme, George Bissias, Michael Fröwis, and the anonymous reviewers for their valuable feedback on earlier versions of this article.

\clearpage

\bibliographystyle{splncs04}
\bibliography{do-not-edit-this-file-manually.bib, do-edit-this-one.bib}

\appendix

\section{Discussion of the Remaining Assumptions}\label{apx:assumptions}

Our work is based on the nine assumptions, \ref{ass:first}--\ref{ass:last}, introduced in Section~\ref{sec:attack_space}.
We have discussed the three most important ones, \ref{ass:attacker}--\ref{ass:daa}, in Section~\ref{sec:attack_space} already, and now cover the rest.

\begin{description}
  \item[To \ref{ass:comm}:]
    We assume reliable (or P2P or Bracha) broadcast, as it is largely orthogonal to consensus.
    Modelling consensus on top of reliable broadcast is common practice~\cite{garay2020SoKConsensus}.
  \item[To \ref{ass:comm} and \ref{ass:dag}:]
    All messages are blocks and these must be organized as a DAG.
    This deliberately restricts our scope.
    In particular, we avoid modelling individual transactions and out-of-band communication.
  \item[To \ref{ass:dag}:]
    We assume a connected DAG with single root block.
    Otherwise we would have to consider permanent splits and could not truncate the common chain.
  \item[To \ref{ass:pow}:]
    We assume that all blocks require the same proof-of-work.
    This is a deliberate restriction to keep the state space tractable.
    Some interesting protocols use more than one puzzle, e.g., Fruitchains~\cite{pass2017FruitChainsFair} and Colordag~\cite{abraham2023ColordagIncentivecompatible}.
    Modelling this is possible by randomly coloring the blocks.
    This however, would add another exponential factor to the state size (all possible colorings).
  \item[To \ref{ass:topo}:]
    We assume all blocks are delivered in topological order.
    Typically protocols set an inductive block validity rule: a block with an invalid parent is invalid itself.
    Hence miners verify the blocks in topological order.
    Mining to confirm unverified blocks is risky.
    Other protocol designs are certainly viable, but we deliberately ignore them.
  \item[To \ref{ass:det}:]
    We assume deterministic behavior for a given BlockDAG and fixed order of delivery.
    This bounds the number of possible protocol-dependent local states: at most one per random transition explored.
    We need this, to explore the state space by traversing all possible state-action-transition paths.
  \item[To \ref{ass:consensus}:]
    We assume the protocol is a consensus protocol in the first place.
    This enables to bound the state space, by enforcing honest behavior and truncating the common chain.
\end{description}

\section{Traditional Selfish Mining Model for Bitcoin}\label{apx:traditional}

\begin{figure}
  \centering
  \pgfdeclarelayer{pattern}
  \pgfdeclarelayer{arrowbg}
  \pgfdeclarelayer{blockbg}
  \pgfdeclarelayer{arrow}
  \pgfsetlayers{pattern, arrowbg, arrow, blockbg, main}
  \tikzstyle{withheld}=[pattern=grid, pattern color=Set1-A!50]
  \tikzstyle{ignored}=[pattern=north west lines, pattern color=Set1-B!60]
  \tikzstyle{visible}=[pattern=north east lines, pattern color=Set1-C!40]
  \begin{tikzpicture}[x=2cm]
    \ifx\theblockid\undefined
      \newcounter{blockid}
    \else
      \setcounter{blockid}{0}
    \fi
    \newcommand{\block}[3]{
      \def\where{#1}
      \def\miner{#2}
      \def\edges{#3}
      \def\label{$\theblockid \mid \miner$}

      \node[draw, fill=white, rounded corners=2pt] at (\where) {\label};
      \begin{pgfonlayer}{blockbg}
        \node[draw, white, ultra thick, rounded corners=2pt] (b\theblockid) at (\where) {\label};
      \end{pgfonlayer}

      \begin{pgfonlayer}{arrow}
        \draw[->, >=stealth] (b\theblockid) \edges;
      \end{pgfonlayer}

      \begin{pgfonlayer}{arrowbg}
        \draw[->, ultra thick, >=stealth, shorten >=-1.5pt, white] (b\theblockid) \edges;
      \end{pgfonlayer}

      \stepcounter{blockid}
    }

    \begin{scope}[drop shadow={fill=white, shadow scale=1.1}]
      \block{0, 0.5}{g}{} %
      \block{1, 1}{d}{edge (b0)} %
      \block{1, 0}{a}{edge (b0)} %
      \block{2, 1}{d}{edge (b1)} %
      \block{2, 0}{a}{edge (b2)} %
      \block{3, 1}{d}{edge (b3)} %
      \block{3, 0}{a}{edge (b4)} %
      \block{4, 0}{a}{edge (b6)} %
    \end{scope}

    \begin{pgfonlayer}{pattern}
      \node[rounded corners, withheld, fit=(b6) (b7)] {};

      \node[rounded corners, ignored, fit=(b1) (b5)] {};

      \node[rounded corners, visible, fit=(b0) (b3) (b4)] {};
    \end{pgfonlayer}
  \end{tikzpicture}

  \tikzstyle{area}=[minimum width=4.3ex, minimum height=2.5ex, rounded corners=3pt]
  \begin{tikzpicture}
    \def\colsep{1em}
    \matrix[row sep=-4pt] {
      \node[right, font=\bfseries] {Blocks:}; &[\colsep]
        \node[]{$g$\strut}; & \node[right] {genesis\strut}; &[\colsep]
          \node[]{$d$\strut}; & \node[right] {defender\strut}; &[\colsep]
            \node[]{$a$\strut}; & \node[right] {attacker\strut}; \\
      \node[right, font=\bfseries] {Masks:}; &[\colsep]
        \node[area, withheld]{}; & \node[right] {withheld\strut}; &[\colsep]
          \node[area, ignored]{}; & \node[right] {ignored\strut}; &[\colsep]
            \node[area, visible]{}; & \node[right] {visible\strut}; \\
      };
  \end{tikzpicture}
  \caption{
    Example state for the Bitcoin protocol.
    Vertices are blocks and arrows indicate parents.
    The available actions are \texttt{Consider(1)}, \texttt{Release(6)}, and \texttt{Continue}.
    Block 5 was just mined; it will become visible during the next \texttt{Continue} action.
    To emulate the traditional \emph{Match} action, the attacker would \texttt{Release(6)} and then \texttt{Continue}.
    To emulate the \emph{Override} action, the attacker would \texttt{Release(6)}, \texttt{Release(7)}, and then \texttt{Continue}.
  }
  \label{fig:state_btc}
\end{figure}
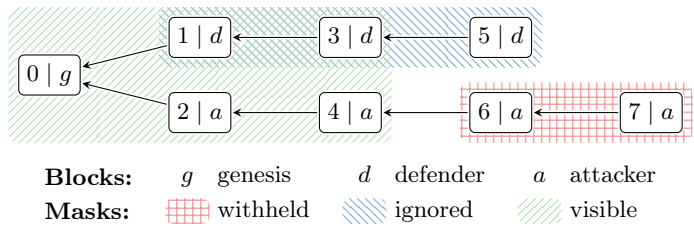

We now present the traditional selfish mining model against Bitcoin~\cite{sapirshtein2016OptimalSelfish}.
Like our attack model (Section~\ref{sec:attack_space}), it features two miners: the attacker and the defender.
The defender extends the longest chain, breaking ties in order first seen.
The attacker can fork the chain, mine blocks in private, and selectively release withheld blocks.
The model is limited to one fork and there are at most two chains: the attacker mines the private chain and the defender mines the public chain.
Figure~\ref{fig:state_btc} shows an example state in \emph{our} generic model, reusing the notation of Figure~\ref{fig:state}.

The traditional model has four actions: \emph{Adopt}, \emph{Match}, \emph{Override}, and \emph{Wait}.
The \emph{Adopt} action discards the private chain. The attacker will start a new fork from the most recent public block.
With \emph{Match}, the attacker releases just enough blocks to induce a block race between the public and the (released subset of the) private chain.
With probability $\gamma$ the block race is resolved in favor of the attacker, that is, the defender discards the public chain and continues mining on the (released subset of the) private chain.
Otherwise, the defender continues mining the public chain.
The \emph{Override} action is similar to \emph{Match}, but the attacker releases one additional block.
This \emph{forces} the defender to discard the public chain with probability one.
Lastly, the \emph{Wait} action lets the attacker continue mining on the private chain without doing anything.
\emph{All actions} induce one block being mined: with probability $\alpha$ the private chain grows by one, otherwise the public chain grows by one.

Note how $\gamma$ models the communication advantage of the attacker.
Honest Bitcoin miners resolve ties in favor of the block first received.
High $\gamma$ implies that the attacker can, in reaction to an honest node mining and sending block~$H$, deliver his own block~$A$ to the remaining defenders before they learn about block~$H$.
These defenders will then discard the public chain in favor of the private chain.
The second parameter, $\alpha$, models the attacker's relative hash rate.

A key feature of this selfish mining model is its simple state space when implemented as MDP:
two non-negative integers describe the lengths of the private and public chain and a ternary variable tracks the feasibility of \emph{Match}.
In the example in Figure~\ref{fig:state_btc}, there are 3 blocks on the public chain, 4 blocks on the private chain, and the \emph{Match} action is feasible.
The remaining information, e.g., block ids and the masks, is redundant.

The downside of such highly optimized MDPs is that they have to be tailored for each protocol individually.
Our approach is to expand the state space such that it covers a wide range of protocols in one go.
We do this, keeping the differences as small as possible:
we do not introduce conflicting assumptions, use the same parameters ($\alpha$ and $\gamma$), and generalize the actions in a protocol-independent way.

\section{Protocol Specifications}%
\label{apx:protocols}

This appendix provides the full protocol specifications that have been omitted in Section~\ref{sec:protocols}.
Listing~\ref{alg:ethereum} specifies Ethereum Proof-of-Work.
Listing~\ref{alg:byzantium} specifies Ethereum's Byzantium upgrade.
Listing~\ref{alg:ghostdag} specifies GhostDAG, with some helper functions deferred to Listing~\ref{alg:ghostdag_util}.
Listing~\ref{alg:parallel} specifies Parallel Proof-of-Work.

\begin{listing}
  \caption{Specification of the Ethereum Protocol}
  \label{alg:ethereum}
  \lstinput{ethereum}
\end{listing}

\begin{listing}
  \caption{Specification of the Byzantium Protocol}
  \label{alg:byzantium}
  \lstinput{byzantium}

  The remaining functionality is inherited from Ethereum as specified in Listing~\ref{alg:ethereum}.
\end{listing}

\begin{listing}
  \caption{Specification of the GhostDAG Protocol}
  \label{alg:ghostdag}
  \lstinput{ghostdag}
\end{listing}

\begin{listing}
  \caption{Specification of the GhostDAG Protocol (Utility Functions)}
  \label{alg:ghostdag_util}
  \lstinput{ghostdag_util}
\end{listing}

\begin{listing}
  \caption{Specification of the Parallel Proof-of-Work Protocol}
  \label{alg:parallel}
  \lstinput{parallel}
\end{listing}

\end{document}